\documentclass[aps,prx,twocolumn,floats,showpacs,superscriptaddress,nofootinbib]{revtex4-1}
\usepackage{amssymb,amsmath,rotate,color,amsfonts}
\usepackage{pifont}
\usepackage[table]{xcolor}
\usepackage{bm}
\usepackage{braket}
\usepackage{tikz}
\usepackage{circuitikz}
\usepackage{float}
\usepackage[caption=false]{subfig}
\usepackage{graphicx}
\usepackage[pagebackref=false,colorlinks,linkcolor=magenta,citecolor=blue,urlcolor=magenta]{hyperref}

\usepackage[title,titletoc,toc]{appendix}
\usepackage{graphicx,epsfig}
\usepackage{times,bbm}

\newcommand{\be}{\begin{equation}}
\newcommand{\ee}{\end{equation}}

\newcommand{\bea}{\begin{eqnarray}}
\newcommand{\eea}{\end{eqnarray}}
\newcommand{\nn}{\nonumber}

\newcommand{\eps}{\varepsilon}

\newcommand{\bk}{{\boldsymbol{k}}}

\newcommand{\br}{{\boldsymbol{r}}}

\newcommand{\bs}{\boldsymbol}
\newcommand{\bmt}{\left[\begin{matrix}}
\newcommand{\emt}{\end{matrix}\right]}

\newcommand{\vect}[1]{\mathbf{#1}}
\newcommand{\Nab}{\bm{\nabla}}

\newcommand{\ii}{\mathrm{i}}

\begin{document}
\preprint{}

\title{
Circuit realization of tilted Dirac cone: A platform for fabrication of curved spacetime geometry on a chip
}
\author{Ahmad Motavassal}
\email{ahmad.motavassel@gmail.com}
\affiliation{Department of Physics$,$ Sharif University of  Technology$,$ Tehran 11155-9161$,$ Iran}
\author{S.A. Jafari}
\email{jafari@sharif.edu}
\affiliation{Department of Physics$,$ Sharif University of  Technology$,$ Tehran 11155-9161$,$ Iran}

\date{\today}

\begin{abstract}
We present a LC circuit model that supports tilted "Dirac cone" in its spectrum. The tilt of the Dirac cone
is specified by the parameters of the model consisting of mutual inductance between the neighboring sites 
and a capacitance $C_0$ at every lattice site. These parameters can be completely measured by impedance spectroscopy. 
Given that a tilted Dirac cone can be described by a background spacetime metric, the impedance spectroscopy can
perfectly provide (local) information about the metric of the spacetime. Non-uniform spatial dependence of the mutual inductance
or capacitance induces non-trivial geometrical structure on the emergent spacetime.
\end{abstract}	
	\maketitle
	
{\em Introduction:}
Dynamics of electrons in solids is shaped by the lattice structure on which they are mounted~\cite{Girvin2019}. 
The constituent electron/ion system can not be separated from the underlying lattice. Circuit electrodynamics
offers an alternative to place circuit elements on complicated lattices. For example the topology of electron bands
of solids can be emulated by circuits~\cite{Ronny2018,Hu2018}. 
But lattices can offer more than band topology:
The first thing that a lattice does is to break the Poincar\'e group~\cite{Ryder1996} into a one of the 230 possible space groups (SGs)~\cite{DresselhausBook}.
Hence the elementary excitations in the solids can be drastically distinct from those in elementary particles physics~\cite{SchwartzBook}.
The irreducible representations of the SG do not allow the band structures to arbitrarily dispers and restricts them by the compatibilitiy relations of
little groups of various high-symmetry points/lines/surfaces~\cite{KittelQTS}. Breaking the Poincar\'e group~\cite{Ryder1996}
also invalidates spin-statistics theorem~\cite{SchwartzBook,Ryder1996} and hence on some lattices fermions may belong to non-spinor 
representation~\cite{Bradlyn2016}, such as spin-1 representation known as triple fermions~\cite{Zhu2016}.
As we will see shortly, the reverse is also possible and a bosonic theory can acquire spinor representation. 

Lattices offer yet another fascinating perspective: It appears that the continuum limit of certain SGs 
corresponds to a spacetime geometry (metric) as detailed below:
A simple nearest neighbor model of fermions on the honeycomb lattice describes Dirac fermions of 
graphene~\cite{Katsnelson2012} that can be interpreted as an emergent Minkowski spacetime. 
It turns out that in certain materials -- notably the $8Pmmn$ borophene that belongs to
SG number $59$ -- the Dirac cone gets tilted~\cite{Zhou2014,Lopez2016,Goerbig2008,Rostamzadeh2019}. The tilting
can be embeded into an emergent metric~\cite{Jafari2019,SaharPolarization,Tohid2019Spacetime,Volovik2016,Volovik2018,Nissinen2017}
$ds^2=-v_F^2dt^2+(d{\bs r}-\bs\zeta v_F dt)^2$, where $v_F$ replaces the speed of light $c$ and is the velocity scale for this emergent spacetime.
In two space dimensions with $\bs\zeta=(\zeta_x,\zeta_y)$, one has
\bea
g_{\mu \nu}=
\begin{bmatrix}
	-1+\zeta^2 & \zeta_x & \zeta_y \\
	\zeta_x & 1 & 0 \\
	\zeta_y & 0 &  1
\end{bmatrix},~~
g^{\mu \nu} =  \begin{bmatrix}
	-1 & -\zeta_x & -\zeta_y \\
	-\zeta_x & 1-\zeta_x^2 & -\zeta_x \zeta_y \\
	-\zeta_y & -\zeta_x \zeta_y & 1-\zeta_y^2
\end{bmatrix}\nn
\label{metric1.eqn}
\eea
where $\zeta^2=\zeta_x^2+\zeta_y^2$ and the above two matrices are inverse of each other~\cite{RyderGR,SchutzGR}. 
$\zeta$ appears as a redshift factor in many quantities, including the density of states~\cite{Mohajerani2021}. 
At $\bs\zeta=0$, the above metric reduces to $\eta_{\mu\nu}={\rm diag}(-1,1,1)$. 

The relation between the {\em space} geometry and certain graphs is well known~\cite{Boettcher2020,Baek2009,Kollar2019}.
Hence it is feasible that dynamics on certain SGs mimics an emergent spacetime. 
The purpose of this paper is to present a LC circuit model on which the dynamics of voltage and
current at long time/distances is governed by the above metric. We will show how the "square root"
of the resulting Klein-Gordon equation is equivalent to a theory of tilted Dirac fermions. 
The same tilted Dirac theory emerges in both electron theory of $8Pmmn$ borophene~\cite{Yasin2021}.
This suggests that the resulting Dirac theory is a property of the underlying lattice.

\begin{figure}[t]
\centering
\includegraphics[clip, trim = 5.5cm 12.7cm 5cm 4.3cm, width=0.28\textwidth]{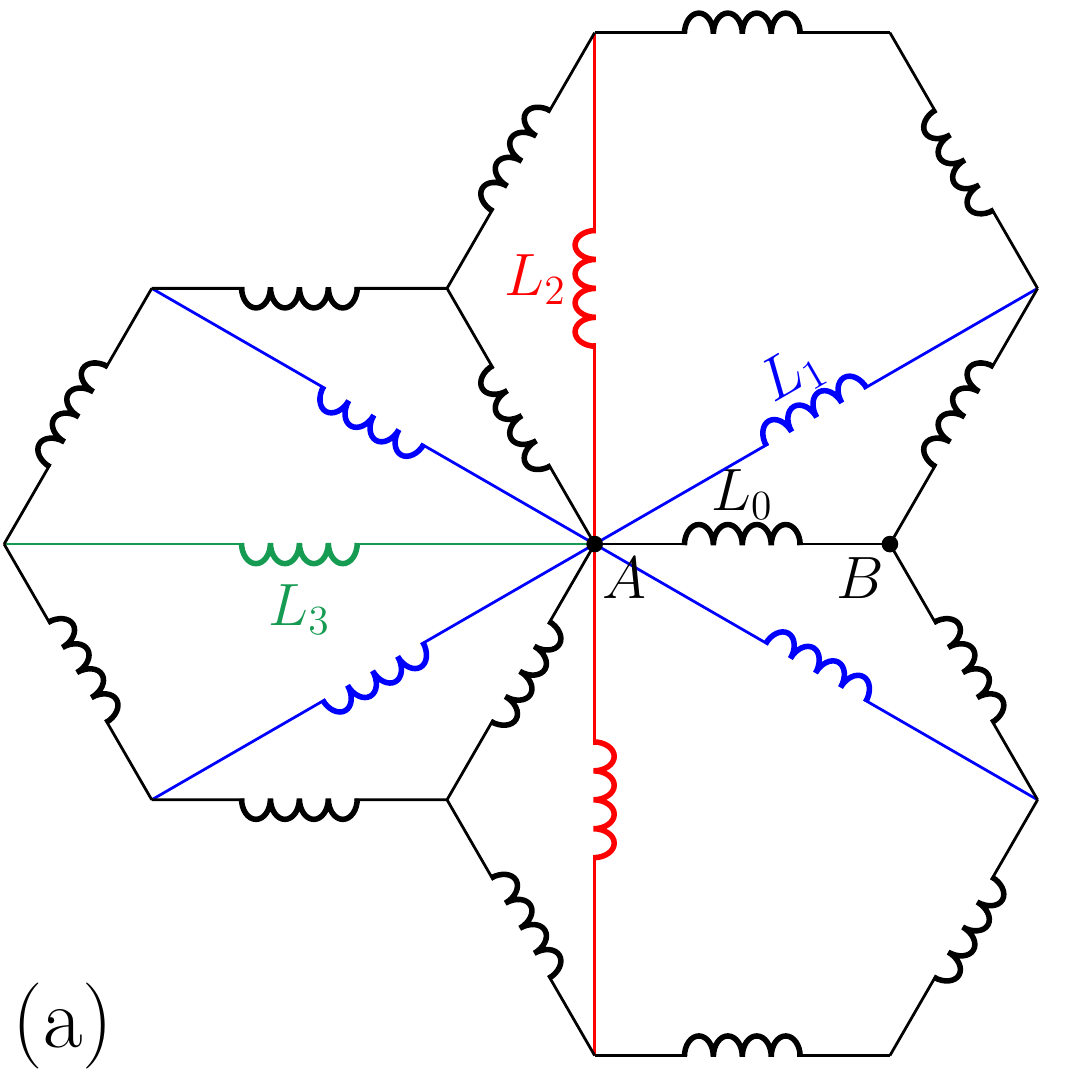}
\includegraphics[clip, trim = 8.2cm 16.6cm 8cm 3cm, width=0.19\textwidth]{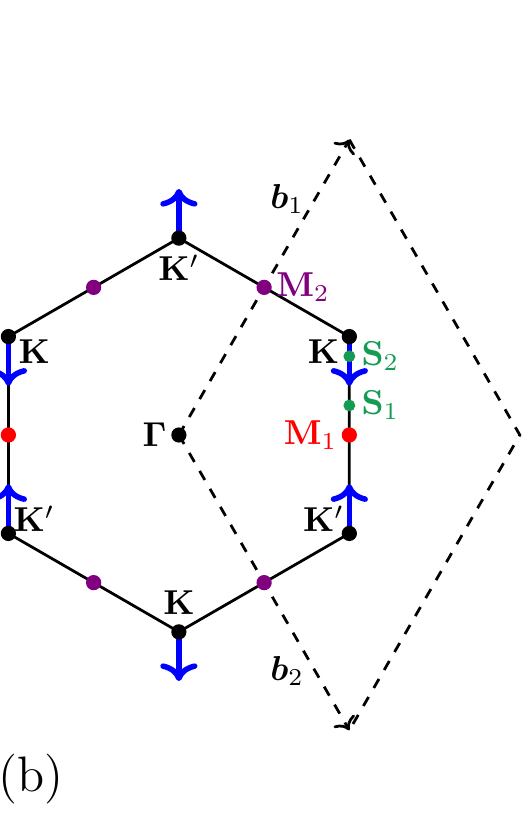}
\caption{(a) The structure of honeycomb circuit. Curly lines of various colors indicate the inductance between various neighbors. For clarity only the neighbors of a single site are drawn. Every node of the lattice is grounded by a capacitance $C_0$. 
(b) Brillouin zone in both Wigner-Seitz (solid honeycomb) and primitive cell (dashed rhombus) constructions.
The reciprocal vectors $\bs{b}_1$ and $\bs{b}_2$ are shown. The blue arrows show the direction of movement of the Dirac nodes 
when we vary $L_3$. The red and purple points label the saddle points $\mathbf{M}_1$ and $\mathbf{M}_2$ (see the text). 
Two other possible exterma denoted by green $\mathbf{S}_1$ and $\mathbf{S}_2$ are verrtically displaced from a Dirac point on both sides.}
\label{fig1}
\end{figure}

{\em Honeycomb lattice circuit model:}
Inspired by our coarse grained~\cite{Kardar2007,Kadanoff2000} fermionic model introduced in Ref.~\cite{Yasin2021}, 
in Fig.~\ref{fig1} we consider a LC circuit based on the periodic honeycomb lattice. Here $L_0$ (black) denotes inductance
between the nearest neighbors. The second neighbor inductances are
of two types, $L_1$ (blue) and $L_2$ (red). The third neighbors along horizontal directon are connected with $L_3$ (green).
Every site is grounded by a capacitance $C_0$. 
The inductance connection enriches the graph structure of a simple honeycomb lattice similar to an effective fermionic hopping model~\cite{Ronny2018},
where the further neighbor connections set the location~\cite{Vozmediano2010} and tilt~\cite{Yasin2021} of the Dirac cone. 
The honeycomb lattice is composed of two Bravis sublattices A and B~\cite{Katsnelson2012}. Setting the length of a bond by $a_0=1/\sqrt 3$, 
the primitive lattice vectors are $\bs{a}_{1|2}=(\sqrt{3},\pm1)/2$. 
Corresponding reciprocal lattice vectors depicted in Fig.~\ref{fig1}(b) are $\bs{b}_{1|2}=2\pi(1,\pm\sqrt{3})/\sqrt 3$.
Suppose that the voltage at at site $\br$ at time $t$ in sublattice A(B) is $V_{1(2)}(\br,t)$.
The Kirchhoff current law for site A reads 
\be
\begin{aligned}
    &\sum_{\bs\delta}\frac{V_1(\br)-V_2(\br+{\bs\delta})}{L_0}+\sum_{i,\lambda}\frac{V_1(\br)-V_1(\br+\lambda {\bs a}_i)}{L_1}\\
    &+\sum_{\lambda}\frac{V_1(\br)-V_1(\br+\lambda({\bs a}_1-{\bs a}_2))}{L_2} \\
    &+\frac{V_1(\br)-V_2(\br-{\bs a}_1-{\bs a}_2)}{L_3}+C_0\frac{d^2V_1(\br)}{dt^2}=0,\nn
\end{aligned}
\label{eq1}
\ee
\noindent
where $\bs\delta$ runs over the three first neighbors, $i=1,2$ labels the basis vectors $\bs a_1$ and $\bs a_2$, $\lambda=\pm$. 
A similar equation for the sublattice B can be written by $\bs\delta\to -\bs\delta$ and $\bs a_{i}\to -\bs a_i$.
Harmonic solutions of the type $V_i(\br,t)=V_i(\br)e^{-\ii\omega t}$ subject to translational invariance
$V_i(\br)=V_i(\bk)e^{\ii\bk.\br}$ give
\bea
\begin{aligned}
  &V_1(\bk)\Big[\nu_0\!-\!
  2a(\cos\bk.{\bs a}_1+\cos\bk.{\bs a}_2)\!-\!2b\cos\bk.({\bs a}_1-{\bs a}_2)\Big]\\
  &-V_2(\bk)\Big[(1+e^{-\ii\bk.{\bs a}_1}+e^{-\ii\bk.{\bs a}_2})-c e^{-\ii\bk.({\bs a}_1+{\bs a}_2)}\Big]\\
  &=\bar\omega^2V_1(\bk),
\end{aligned}\nn
  \label{V1kA.eqn}
\eea
where we have defined dimensionless (and positive) parameters $a=\frac{L_0}{L_1}$, $b=\frac{L_0}{L_2}$, $c=\frac{L_0}{L_3}$  and $\nu_0= 3+4a+2b+c$.
The frequency $\omega_*^2=(L_0C_0)^{-1}$ is the natural frequency of the system that allows to define dimensionless frequency $\bar\omega$ by $\omega=\omega_*\bar\omega$. 
Putting together the equations for A and B results in the eigenvalue problem for the {\em Dynamical matrix} $D(\bs k)$,
\begin{equation}
\begin{pmatrix}
\eps(\bk)  & \Delta(\bk)  \\
\Delta^{\ast}(\bk) & \eps(\bk) 
\end{pmatrix}
\begin{pmatrix}
V_1(\bk) \\
V_2(\bk)
\end{pmatrix}
=\bar\omega^2({\bk})
\begin{pmatrix}
V_1(\bk) \\
V_2(\bk)
\end{pmatrix}
\label{eigen.eqn}
\end{equation}
where $\eps(\bs k) =  \nu_0-4a \cos k_x \frac{\sqrt{3}}{2} \cos \frac{k_y}{2} -2b \cos k_y$ and 
$\Delta(\bs k) =  |\Delta|e^{\ii\phi} = -1-2e^{-\ii k_x \frac{\sqrt{3}}{2}} \cos \frac{k_y}{2} -c e^{-\ii k_x\sqrt{3}}$.
Despite that quantization of the current-voltage oscillator gives a bosonic theory, 
a spinor structure naturally emerges from the two-sublattice nature of the  honeycomb lattice
$\langle V_{\bs k}|=
\begin{pmatrix} V_1(\bk) & V_2(\bk)  \end{pmatrix}$. 
Eq.~\eqref{eigen.eqn} gives $\bar\omega^{2}_\pm(\bs k)=\eps(\bs k)\pm|\Delta(\bs k)|$
and  $\ket{V_{\pm,\bs k})}=\frac{1}{\sqrt{2}} \begin{pmatrix} \pm e^{\ii\phi} \\ 1 \end{pmatrix} $, where $\phi$ is 
the phase of the complex number $\Delta$. 
The splitting between the upper ($+$) and lower ($-$) frequency bands is controlled by $\Delta(\bs k)$ where
\be
\begin{aligned}
&|\Delta(\bs k)|^2=4\Big[\cos^2{\frac{k_y}{2}}+(1+c)\cos{\frac{k_y}{2}}\cos{\frac{k_x\sqrt{3}}{2}}\\
&+c\cos^2{\frac{k_x\sqrt{3}}{2}}+\left(\frac{1-c}{2}\right)^2\Big].
\end{aligned}
\label{absdelta.eqn}
\ee
The upper and lower bands meet when, $0\leq c \leq 1$. 
When $c=0$ gap closing (nodes) are located on the corners of BZ as shown in Fig.~\ref{fig1}(b) in Wigner-Seitz and primitive cell representations. 
The coordinates of $\mathbf{K}/\mathbf{K}'$ are  $(0,\mp 4\pi/3)$.
Upon increasing $c$, the horizental coordinate of these points do not change, but because of the increase in $\cos \frac{k_y}{2}$, 
$\mathbf{K}$ and $\mathbf{K}'$ move vertically towards each other.  
Hence the role of parameter $c$ is to control the location of the two independent (see the rhombos primitive cell BZ) nodes.
Increasing $c$ from $0$ to $1$ \textit{shifts} the two nodes toward each other. 
At $c=1$ these two points collide and anihilate at the $M_1$ point -- due to their opposite topological charge --
giving fully gapped spectrum for $c>1$. 

{\em Dirac theory:}
$\eps(\bs k)$ and $\Delta(\bs k)$ near the gap closing become
\bea
\begin{aligned}
   &\eps(\bs k) \approx  \bar\omega_0^2+ \tau \alpha_y \delta k_y,~~
     \Delta \approx  -\ii w_x \delta k_x +\tau w_y \delta k_y,\\
   &\bar\omega_0^2 =  (3+c)(1+2a+b(1-c)) 
\end{aligned}
\label{opf.eqn}
\eea
where $\tau=\pm$ marks the node (valley) around which the linearization has been made,
$(\delta k_x, \delta k_y)$ are the deviations from the gap closing point, and $\alpha_y =  \sqrt{(1-c)(3+c)}(a-b(1+c))$
while $ w_x =  \sqrt{3}(1-c)/2$ and $w_y =  \sqrt{(1-c)(3+c)}/2$ that determine the dynamical matrix~\eqref{eigen.eqn}.
By Taylor expansion of the square root of the matrix $D(\bs k)$ around $\bar\omega^2_0$, one obtains a new matrix $h(\bs k)$ whose eigenvalues are $\bar\omega(\bs k)$:
\bea
\begin{aligned}
   &h(\bs k) =(\bar\omega_0+ \bs v_t \cdot \delta\bs k)\sigma_0 + (-\ii v_x\delta k_x\sigma_x+\tau v_y\delta k_y\sigma_y), \\
   &v_x=\frac{w_x}{2\bar\omega_0},~
   v_y=\frac{w_y}{2\bar\omega_0},~
   v_{ty}=\frac{\alpha_y}{2\bar\omega_0},~
   v_{tx}=0,
\end{aligned}
\label{tDirac.eqn}
\eea
where $\bs v_t$ is the "tilt" velocity scale that defiens the tilt parameter by $\zeta_a=v_{ta}/v_a, a=x,y$. 
Therefore, close the operation frequency $\bar\omega_0$ given by Eq.~\eqref{opf.eqn}, 
the matrix whose eigenvalues give the eigen-frequencies of our circuit system are given in Eq.~\eqref{tDirac.eqn}
that describes tilted Dirac fermions. The above tilted Dirac theory can be regarded as the "square root" 
of theory described by $D(\bs k)$, the same way that Dirac equation is regarded as the square root of 
Klein-Gordon equation~\cite{Ryder1996}.  

\begin{table}[b]
\begin{ruledtabular}
\begin{tabular}{l|l|l|l}
 & $\vect{k}$ & $\Omega_\pm$ & type \\ \hline
$\mathbf{\Gamma}$ & $(0,0)$ & $0$, $6+2c$ & Max/Min \\
$\mathbf{M}_1$ & $(\pm \frac{2\pi}{\sqrt{3}},0)$ & $2+8a+2c$, $4+8a$ & Saddle \\
$\mathbf{M}_2$ & $(\pm \frac{\pi}{\sqrt{3}}, \pm \pi)$ & $2+4a+4b+2c$, & Saddle \\
               &                                       & $4+4a+4b$     &  \\
$\mathbf{S}_1$ & $(\pm\frac{2\pi}{\sqrt{3}},2\cos^{-1}(\mp \frac{1+2a}{4b}))$ & $\frac{1}{4b}(1+2a+4b)^2$ & Max/Min \\ 
$\mathbf{S}_2$ & $(\pm\frac{2\pi}{\sqrt{3}},2\cos^{-1}(\mp \frac{1-2a}{4b}))$ & $2c+6+$ & Max/Min \\ 
               &                                       & $\frac{1}{4b}(-1+2a+4b)^2$ &  \\ 
\end{tabular}
\end{ruledtabular}
\caption{Extrma and their corresponding values of $\Omega$.}
\label{table1}
\end{table}

{\em Spectral density:} For the rest of this work we do not need $h(\bs k)$ and continue to work with 
$D(\bs k)$. So we define a new symbol $\Omega(\bs k)=\bar\omega^2(\bs k)$ to label its eigenvalues. This is because the impedance spectroscopy
will directly measure the spectrum of the $D(\bs k)$, not the Dirac Hamiltonian~\eqref{tDirac.eqn}. 
The resolvant~\cite{economou_2006} of the $D(\bs k)$ that describes the dynamics of voltage/current on the graph is
\begin{equation}
G(\bs k,z) = \frac{1}{2} \frac{1}{(z-\eps(\bs k))^2-|\Delta(\bs k)|^2}
\begin{pmatrix}
z-\eps(\bs k) & \Delta(\bs k) \\
\Delta^\ast(\bs k) & z-\eps(\bs k),\nn
\end{pmatrix}
\end{equation}
the imaginary part of which is defined by $\rho = -\frac{1}{\pi} \Im\{\mbox{tr}[G^+]\}$
where $G^+(\bs k,\lambda) = G(\bs k,\lambda+\ii 0^+)$ gives the density of states (DOS). 
The trace includes summation over the diagonal elements of $G$ and  integration over the whole BZ.

When the spectral density is plotted as a function  of $\Omega$, contains a great deal of information.
The first imortant feature of the density of $\Omega$ values is the location of the Dirac node, 
Eq.~\eqref{opf.eqn}, $\bar\omega^2_0=\Omega_0$ that solves $\Omega_+(\bs k)=\Omega_-(\bs k)$. 
This gives the first relation 
among the model parameters $a,b,c$ that can be directly read off from the DOS. 
The exterma of DOS are determined from $\Nab_{\bs k}\Omega(\bs k)=0$. 
In Tab.~\ref{table1} we list the positions and values of $\Omega$ at two van-Hove singularities
$\mathbf{M}_1$ and $\mathbf{M}_2$ shown in Fig.~\ref{fig1}(b). 
Because $\mathbf{M}_1$ and $\mathbf{M}_2$ are saddle points, they give logarithmic van Hove singularities
the locations of which directly relate to model parameters as,
\bea
&&(\Omega_+-\Omega_-)|_{\mathbf{M}_1}=(\Omega_+-\Omega_-)|_{\mathbf{M}_2}=2(1-c),\\
&&\Omega_+|_{\mathbf{M}_2}-\Omega_+|_{\mathbf{M}_1}=\Omega_-|_{\mathbf{M}_2}-\Omega_-|_{\mathbf{M}_1}=4(b-a).
\eea
The first equation tells us that the van Hove singularities arising from a given point in upper and lower branches are 
separated by deviations of $c$ from $1$. This helps to immediately read off the parameter $c$. 
The second equation above implies that the separation of van Hove singularities in upper branch is controlled by $b-a$ and 
when $a=b$ the van Hove singularities for $\mathbf{M}_1$ and $\mathbf{M}_2$ points coincide and hence the number of van Hove singularities is
reduced by two. Now let us see how one can measure the location of the above singularities. 

\begin{figure}[t]
	\centering
	\includegraphics[clip, trim=4.9cm 9.2cm 4.6cm 9cm, width=0.45\textwidth]{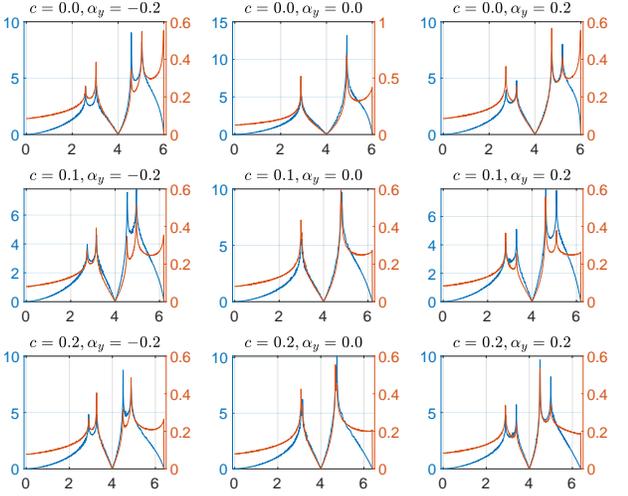}
	\caption{The impedance and DOS curves vs. $\Omega$ for differnt values of $c$ and $\alpha_y$. The Dirac level $\Omega_0$ is held fixed at $4$. 
	The blue (oragne) curve with axis range on left (right) is $\operatorname{Re}\{Z_{AA}(\bs a_1)\}$ (DOS).}
	\label{fig2}
\end{figure}

\begin{table}[b]
\begin{ruledtabular}
\begin{tabular}{l|c|c}
Parity of $(m,n)$ & $\mathbf{M}_1$ & $\mathbf{M}_2$ \\ \hline
(even, even) & \ding{55} & \ding{55} \\
(odd, odd) & \ding{55} & \ding{51} \\
(even, odd) or (odd, even) & \ding{51} & \ding{51}
\end{tabular}
\end{ruledtabular}
\caption{How to select DOS singularities in impedance spectroscopy. 
\ding{51}/\ding{55} indicate the presence/absence of the DOS singularity.  }
\label{table2}
\end{table}

{\em Impedance spectroscopy:}
This measurement consists in sending a current through one node into our LC lattice and extracting the
current through another (arbitrary) node and corresponds to adding a non-zero current to the right side of Eq.~\eqref{eq1}~\cite{Cserti2011}. 
The operation frequency can be adjusted at will to probe the Dirac physics near the crossing point $\Omega_0$. 
If the current is sent in to site $\bs r_a$ on sublattice $\nu_1$ and is extracted from site $\bs r_b$ on sublattice $\nu_2$,
then one has to add 
$ I_\nu(\bs r)=I_0(\delta_{\nu\nu_1}\delta(\bs r-\bs r_a)-\delta_{\nu\nu_2}\delta(\bs r-\bs r_b))$ or
in Fourier representation 
$I_\nu(\bs k)=I_0(\delta_{\nu\nu_1}e^{-\ii\bs k.\bs{r}_a}-\delta_{\nu\nu_2}e^{-\ii\bs{k}.\bs{r}_b})$
to the right side of Eq.~\eqref{eigen.eqn} that gives,
$[D(\bs{k})-\Omega\sigma_0]\ket{V}=-\ii\omega L_0 \ket{ I(\bs {k})}$. By 
$D(\bs{k})-\Omega\sigma_0=-G^{-1}(\bs{k},\Omega)$,
we obtain the matrix equation $V_\mu=\ii\omega L_0G_{\mu\nu}I_\nu$ where $\mu,\nu=1,2$ label the sublattices. 
By definition of impedance (the difference between the voltages of the nodes devided by the current) we get
\bea
  Z_{\nu_1\nu_2}(\bs{r}_a-\bs{r}_b)&=&\ii\omega_* L_0\sqrt{\Omega}\frac{1}{N}\sum_{\bs k} \Big[ G^{}_{\nu_1\nu_1}+G_{\nu_2\nu_2}\label{imp.eqn}\\
   &&-G_{\nu_1\nu_2}e^{\ii\bs{k}.(\bs{r}_a-\bs{r}_b)}-G_{\nu_2\nu_1}e^{-\ii\bs{k}.(\bs{r}_a-\bs{r}_b)}\Big],\nn
\eea
where $N$ is the number of unit cells.
The frequency dependence is implied for the Green's function matrix elements $G_{\nu_i\nu_j}$. 
For large enough lattices with many degrees of freedom, the sum over $\bs {k}$ can be replaced by an integral over BZ. 
This completes the expression of the impedance in terms of the Green's function. It further suggests to work with the "normalized impedance" 
$Z_{\nu\nu}/(\omega_* L_0)$. As can be seen in Fig.~\ref{fig2}, the (local) impedance measured between a typical in-out points separated by $\bs a_1$
clearly contains information about he essential features of the DOS and hence serves as a spectroscopic determination tool to measure $a,b,c$ parameters. 

\begin{figure}[t]
\centering
\includegraphics[clip, trim=5cm 9.2cm 4.7cm 9.1cm, width=0.45\textwidth]{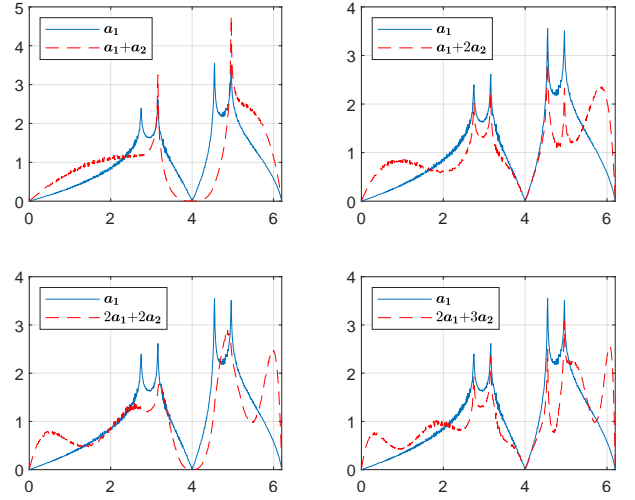}
\caption{The comparison between the impedance $\operatorname{Re}\{Z_{AA}(\bs r)\}$ at $\bar\omega_0^2=4$, $c=0.1$, $\alpha_y=-0.2$
for $\bs r=\bs a_1$ (blue solid curve) and  $\bs r=m\bs a_1+n\bs a_2$ (red dashed curve).  }
\label{fig3}
\end{figure}
If one probes the {\em non-local} impedance between arbitray unit cells separated by 
$\bs{r}_a-\bs{r}_b=m\bs a_1+n\bs a_2=((m+n)\sqrt{3}/2,(m-n)/2)$, the diagonal 
component of Eq.~\eqref{imp.eqn} gives
\begin{equation}
\sum_{\bs k}2G^{}_{\nu_1\nu_1}(1-\cos{(\bs{k}.(\bs{r}_a-\bs{r}_b))})=\sum_{\bs k}2G^{}_{\nu_1\nu_1}\Delta\phi. 
\end{equation}	
The term in the paranthesis denoted by $\Delta\phi$ resembles the atomic interference term that
arises in the scattering determination of crystal structure~\cite{Ibach2009}. For $\bs k_{\mathbf{M}_1}=(\pm\frac{2\pi}{\sqrt{3}},0)$ and $\bs k_{\mathbf{M}_2}=(\pm\frac{\pi}{\sqrt{3}},\pm\pi)$ the interference terms become
\begin{eqnarray}
&\Delta\phi_{\mathbf{M}_1}=1-\cos((m+n)\pi)=1-(-1)^{m+n}, \nn \\
&\Delta\phi_{\mathbf{M}_2} = \left\{
\begin{array}{ll}
1-\cos{m\pi} & \text{same sign}\\
1-\cos{n\pi} & \text{opposite sign}
\end{array} \right. = \left\{
\begin{array}{l}
1-(-1)^m\\
1-(-1)^n
\end{array} \right., \nn
\end{eqnarray}	
which shows that when $m+n$ is even, the van-Hove singularity at $\mathbf{M}_1$ disappears by interference. In order to annihilate the $\mathbf{M}_2$ singularity, both $m$ and $n$ must be even. Tab.\ref{table2} summarizes the above interference physics of van-Hove singularities. Fig.~\ref{fig3} compares the impedance bewtween points separated by $\bs a_1$ that contains full DOS singularities, with
few other $Z$ for various $m\bs a_1+n\bs a_2$ values, in agreement with Tab.~\ref{table2}.  

{\em Outlook:}
We have presented a honeycomb lattice model for circuit realization of a tilted "Dirac cone"
and its local and non-local impedance spectroscopy to fully determine the model parameters. Allowing the model parameters to vary in on the lattice
will imprint a spacetime geometry that can be arbitrarily tuned. 
Our model is a step towards "on chip" realization of interesting {\em spacetime} geometries. 
The "particles" in this system are current/voltage pulses that can be traced 
by appropriate impedance spectroscopy whose line shap contains complete {\em local} information about the
parameter of the model, and hence the properties of the spacetime that emerges at long distances.
Our current study shows that the relation between the space group and the
ensuing spacetime geometry at long distances is the same for $2p$ electrons in $8Pmmn$ borophene, and current pulses. 

The connection between graphs and {\em space} geometry in the context of circuit electrodynamics~\cite{Gorshkov2020}
as well as in the band theory~\cite{Maciejko2021} and possible implications for high-temperateure superconductivity
has been discussed~\cite{Campi2015}. 
Our proposal differs in that it offers a wider perspective for the fabrication and manipulation of {\em spacetime} geometry~\cite{Tohid2020Synthetic},
not merely the space geometry~\cite{Boettcher2020,Baek2009,Kollar2019}. As such, our setup allows for emulation of various "gravitational" phenomena.
When one is dealing with a pure space geometry, the effect of curvature can be replaced by a pseudo $U(1)$ gauge field,
while in the case of spacetime geomery, one requires non-Abelian gauge fields. 
As such, our circuit model can be regarded as a convenient platform for the synthesis of non-Abelian guage fields~\cite{Tohid2020Synthetic}.

\bibliography{mybib.bib}

\begin{thebibliography}{37}%
\makeatletter
\providecommand \@ifxundefined [1]{%
 \@ifx{#1\undefined}
}%
\providecommand \@ifnum [1]{%
 \ifnum #1\expandafter \@firstoftwo
 \else \expandafter \@secondoftwo
 \fi
}%
\providecommand \@ifx [1]{%
 \ifx #1\expandafter \@firstoftwo
 \else \expandafter \@secondoftwo
 \fi
}%
\providecommand \natexlab [1]{#1}%
\providecommand \enquote  [1]{``#1''}%
\providecommand \bibnamefont  [1]{#1}%
\providecommand \bibfnamefont [1]{#1}%
\providecommand \citenamefont [1]{#1}%
\providecommand \href@noop [0]{\@secondoftwo}%
\providecommand \href [0]{\begingroup \@sanitize@url \@href}%
\providecommand \@href[1]{\@@startlink{#1}\@@href}%
\providecommand \@@href[1]{\endgroup#1\@@endlink}%
\providecommand \@sanitize@url [0]{\catcode `\\12\catcode `\$12\catcode
  `\&12\catcode `\#12\catcode `\^12\catcode `\_12\catcode `\%12\relax}%
\providecommand \@@startlink[1]{}%
\providecommand \@@endlink[0]{}%
\providecommand \url  [0]{\begingroup\@sanitize@url \@url }%
\providecommand \@url [1]{\endgroup\@href {#1}{\urlprefix }}%
\providecommand \urlprefix  [0]{URL }%
\providecommand \Eprint [0]{\href }%
\providecommand \doibase [0]{http://dx.doi.org/}%
\providecommand \selectlanguage [0]{\@gobble}%
\providecommand \bibinfo  [0]{\@secondoftwo}%
\providecommand \bibfield  [0]{\@secondoftwo}%
\providecommand \translation [1]{[#1]}%
\providecommand \BibitemOpen [0]{}%
\providecommand \bibitemStop [0]{}%
\providecommand \bibitemNoStop [0]{.\EOS\space}%
\providecommand \EOS [0]{\spacefactor3000\relax}%
\providecommand \BibitemShut  [1]{\csname bibitem#1\endcsname}%
\let\auto@bib@innerbib\@empty
\bibitem [{\citenamefont {Girvin}\ and\ \citenamefont
  {Yang}(2019)}]{Girvin2019}%
  \BibitemOpen
  \bibfield  {author} {\bibinfo {author} {\bibfnamefont {S.~M.}\ \bibnamefont
  {Girvin}}\ and\ \bibinfo {author} {\bibfnamefont {K.}~\bibnamefont {Yang}},\
  }\href {\doibase 10.1017/9781316480649} {\emph {\bibinfo {title} {Modern
  Condensed Matter Physics}}}\ (\bibinfo  {publisher} {Cambridge University
  Press},\ \bibinfo {year} {2019})\BibitemShut {NoStop}%
\bibitem [{\citenamefont {Lee}\ \emph {et~al.}(2018)\citenamefont {Lee},
  \citenamefont {Imhof}, \citenamefont {Berger}, \citenamefont {Bayer},
  \citenamefont {Brehm}, \citenamefont {Molenkamp}, \citenamefont {Kiessling},\
  and\ \citenamefont {Thomale}}]{Ronny2018}%
  \BibitemOpen
  \bibfield  {author} {\bibinfo {author} {\bibfnamefont {C.~H.}\ \bibnamefont
  {Lee}}, \bibinfo {author} {\bibfnamefont {S.}~\bibnamefont {Imhof}}, \bibinfo
  {author} {\bibfnamefont {C.}~\bibnamefont {Berger}}, \bibinfo {author}
  {\bibfnamefont {F.}~\bibnamefont {Bayer}}, \bibinfo {author} {\bibfnamefont
  {J.}~\bibnamefont {Brehm}}, \bibinfo {author} {\bibfnamefont {L.~W.}\
  \bibnamefont {Molenkamp}}, \bibinfo {author} {\bibfnamefont {T.}~\bibnamefont
  {Kiessling}}, \ and\ \bibinfo {author} {\bibfnamefont {R.}~\bibnamefont
  {Thomale}},\ }\href {\doibase 10.1038/s42005-018-0035-2} {\bibfield
  {journal} {\bibinfo  {journal} {Communications Physics}\ }\textbf {\bibinfo
  {volume} {1}} (\bibinfo {year} {2018}),\
  10.1038/s42005-018-0035-2}\BibitemShut {NoStop}%
\bibitem [{\citenamefont {Li}\ \emph {et~al.}(2018)\citenamefont {Li},
  \citenamefont {Sun}, \citenamefont {Zhu}, \citenamefont {Guo}, \citenamefont
  {Jiang}, \citenamefont {Kariyado}, \citenamefont {Chen},\ and\ \citenamefont
  {Hu}}]{Hu2018}%
  \BibitemOpen
  \bibfield  {author} {\bibinfo {author} {\bibfnamefont {Y.}~\bibnamefont
  {Li}}, \bibinfo {author} {\bibfnamefont {Y.}~\bibnamefont {Sun}}, \bibinfo
  {author} {\bibfnamefont {W.}~\bibnamefont {Zhu}}, \bibinfo {author}
  {\bibfnamefont {Z.}~\bibnamefont {Guo}}, \bibinfo {author} {\bibfnamefont
  {J.}~\bibnamefont {Jiang}}, \bibinfo {author} {\bibfnamefont
  {T.}~\bibnamefont {Kariyado}}, \bibinfo {author} {\bibfnamefont
  {H.}~\bibnamefont {Chen}}, \ and\ \bibinfo {author} {\bibfnamefont
  {X.}~\bibnamefont {Hu}},\ }\href {\doibase 10.1038/s41467-018-07084-2}
  {\bibfield  {journal} {\bibinfo  {journal} {Nature Communications}\ }\textbf
  {\bibinfo {volume} {9}} (\bibinfo {year} {2018}),\
  10.1038/s41467-018-07084-2}\BibitemShut {NoStop}%
\bibitem [{\citenamefont {Ryder}(1996)}]{Ryder1996}%
  \BibitemOpen
  \bibfield  {author} {\bibinfo {author} {\bibfnamefont {L.~H.}\ \bibnamefont
  {Ryder}},\ }\href {\doibase 10.1017/cbo9780511813900} {\emph {\bibinfo
  {title} {Quantum Field Theory}}}\ (\bibinfo  {publisher} {Cambridge
  University Press},\ \bibinfo {year} {1996})\BibitemShut {NoStop}%
\bibitem [{\citenamefont {Dresselhaus}(2008)}]{DresselhausBook}%
  \BibitemOpen
  \bibfield  {author} {\bibinfo {author} {\bibfnamefont {M.~S.}\ \bibnamefont
  {Dresselhaus}},\ }\href@noop {} {\emph {\bibinfo {title} {Group Theory:
  Applications to the Physics of Condensed Matter}}}\ (\bibinfo  {publisher}
  {Springer},\ \bibinfo {year} {2008})\BibitemShut {NoStop}%
\bibitem [{\citenamefont {Schwartz}(2014)}]{SchwartzBook}%
  \BibitemOpen
  \bibfield  {author} {\bibinfo {author} {\bibfnamefont {M.~D.}\ \bibnamefont
  {Schwartz}},\ }\href@noop {} {\emph {\bibinfo {title} {Quantum Field Theory
  and the Standard Model}}}\ (\bibinfo  {publisher} {Cambridge University
  Press},\ \bibinfo {year} {2014})\BibitemShut {NoStop}%
\bibitem [{\citenamefont {Kittel}(1987)}]{KittelQTS}%
  \BibitemOpen
  \bibfield  {author} {\bibinfo {author} {\bibfnamefont {C.}~\bibnamefont
  {Kittel}},\ }\href@noop {} {\emph {\bibinfo {title} {Quantum Theory of
  Solids}}}\ (\bibinfo  {publisher} {John Wiley \& Sons},\ \bibinfo {year}
  {1987})\BibitemShut {NoStop}%
\bibitem [{\citenamefont {Bradlyn}\ \emph {et~al.}(2016)\citenamefont
  {Bradlyn}, \citenamefont {Cano}, \citenamefont {Wang}, \citenamefont
  {Vergniory}, \citenamefont {Felser}, \citenamefont {Cava},\ and\
  \citenamefont {Bernevig}}]{Bradlyn2016}%
  \BibitemOpen
  \bibfield  {author} {\bibinfo {author} {\bibfnamefont {B.}~\bibnamefont
  {Bradlyn}}, \bibinfo {author} {\bibfnamefont {J.}~\bibnamefont {Cano}},
  \bibinfo {author} {\bibfnamefont {Z.}~\bibnamefont {Wang}}, \bibinfo {author}
  {\bibfnamefont {M.~G.}\ \bibnamefont {Vergniory}}, \bibinfo {author}
  {\bibfnamefont {C.}~\bibnamefont {Felser}}, \bibinfo {author} {\bibfnamefont
  {R.~J.}\ \bibnamefont {Cava}}, \ and\ \bibinfo {author} {\bibfnamefont
  {B.~A.}\ \bibnamefont {Bernevig}},\ }\href {\doibase 10.1126/science.aaf5037}
  {\bibfield  {journal} {\bibinfo  {journal} {Science}\ }\textbf {\bibinfo
  {volume} {353}},\ \bibinfo {pages} {aaf5037} (\bibinfo {year}
  {2016})}\BibitemShut {NoStop}%
\bibitem [{\citenamefont {Zhu}\ \emph {et~al.}(2016)\citenamefont {Zhu},
  \citenamefont {Winkler}, \citenamefont {Wu}, \citenamefont {Li},\ and\
  \citenamefont {Soluyanov}}]{Zhu2016}%
  \BibitemOpen
  \bibfield  {author} {\bibinfo {author} {\bibfnamefont {Z.}~\bibnamefont
  {Zhu}}, \bibinfo {author} {\bibfnamefont {G.~W.}\ \bibnamefont {Winkler}},
  \bibinfo {author} {\bibfnamefont {Q.}~\bibnamefont {Wu}}, \bibinfo {author}
  {\bibfnamefont {J.}~\bibnamefont {Li}}, \ and\ \bibinfo {author}
  {\bibfnamefont {A.~A.}\ \bibnamefont {Soluyanov}},\ }\href {\doibase
  10.1103/physrevx.6.031003} {\bibfield  {journal} {\bibinfo  {journal}
  {Physical Review X}\ }\textbf {\bibinfo {volume} {6}} (\bibinfo {year}
  {2016}),\ 10.1103/physrevx.6.031003}\BibitemShut {NoStop}%
\bibitem [{\citenamefont {Katsnelson}(2012)}]{Katsnelson2012}%
  \BibitemOpen
  \bibfield  {author} {\bibinfo {author} {\bibfnamefont {M.~I.}\ \bibnamefont
  {Katsnelson}},\ }\href {\doibase 10.1017/cbo9781139031080} {\emph {\bibinfo
  {title} {Graphene}}}\ (\bibinfo  {publisher} {Cambridge University Press},\
  \bibinfo {year} {2012})\BibitemShut {NoStop}%
\bibitem [{\citenamefont {Zhou}\ \emph {et~al.}(2014)\citenamefont {Zhou},
  \citenamefont {Dong}, \citenamefont {Oganov}, \citenamefont {Zhu},
  \citenamefont {Tian},\ and\ \citenamefont {Wang}}]{Zhou2014}%
  \BibitemOpen
  \bibfield  {author} {\bibinfo {author} {\bibfnamefont {X.-F.}\ \bibnamefont
  {Zhou}}, \bibinfo {author} {\bibfnamefont {X.}~\bibnamefont {Dong}}, \bibinfo
  {author} {\bibfnamefont {A.~R.}\ \bibnamefont {Oganov}}, \bibinfo {author}
  {\bibfnamefont {Q.}~\bibnamefont {Zhu}}, \bibinfo {author} {\bibfnamefont
  {Y.}~\bibnamefont {Tian}}, \ and\ \bibinfo {author} {\bibfnamefont {H.-T.}\
  \bibnamefont {Wang}},\ }\href {\doibase 10.1103/PhysRevLett.112.085502}
  {\bibfield  {journal} {\bibinfo  {journal} {Phys. Rev. Lett.}\ }\textbf
  {\bibinfo {volume} {112}},\ \bibinfo {pages} {085502} (\bibinfo {year}
  {2014})}\BibitemShut {NoStop}%
\bibitem [{\citenamefont {Lopez-Bezanilla}\ and\ \citenamefont
  {Littlewood}(2016)}]{Lopez2016}%
  \BibitemOpen
  \bibfield  {author} {\bibinfo {author} {\bibfnamefont {A.}~\bibnamefont
  {Lopez-Bezanilla}}\ and\ \bibinfo {author} {\bibfnamefont {P.~B.}\
  \bibnamefont {Littlewood}},\ }\href {\doibase 10.1103/PhysRevB.93.241405}
  {\bibfield  {journal} {\bibinfo  {journal} {Phys. Rev. B}\ }\textbf {\bibinfo
  {volume} {93}},\ \bibinfo {pages} {241405} (\bibinfo {year}
  {2016})}\BibitemShut {NoStop}%
\bibitem [{\citenamefont {Goerbig}\ \emph {et~al.}(2008)\citenamefont
  {Goerbig}, \citenamefont {Fuchs}, \citenamefont {Montambaux},\ and\
  \citenamefont {Pi\'echon}}]{Goerbig2008}%
  \BibitemOpen
  \bibfield  {author} {\bibinfo {author} {\bibfnamefont {M.~O.}\ \bibnamefont
  {Goerbig}}, \bibinfo {author} {\bibfnamefont {J.-N.}\ \bibnamefont {Fuchs}},
  \bibinfo {author} {\bibfnamefont {G.}~\bibnamefont {Montambaux}}, \ and\
  \bibinfo {author} {\bibfnamefont {F.}~\bibnamefont {Pi\'echon}},\ }\href
  {\doibase 10.1103/PhysRevB.78.045415} {\bibfield  {journal} {\bibinfo
  {journal} {Phys. Rev. B}\ }\textbf {\bibinfo {volume} {78}},\ \bibinfo
  {pages} {045415} (\bibinfo {year} {2008})}\BibitemShut {NoStop}%
\bibitem [{\citenamefont {Rostamzadeh}\ \emph {et~al.}(2019)\citenamefont
  {Rostamzadeh}, \citenamefont {Adagideli},\ and\ \citenamefont
  {Goerbig}}]{Rostamzadeh2019}%
  \BibitemOpen
  \bibfield  {author} {\bibinfo {author} {\bibfnamefont {S.}~\bibnamefont
  {Rostamzadeh}}, \bibinfo {author} {\bibfnamefont {i.~d. I. m.~c.}\
  \bibnamefont {Adagideli}}, \ and\ \bibinfo {author} {\bibfnamefont {M.~O.}\
  \bibnamefont {Goerbig}},\ }\href {\doibase 10.1103/PhysRevB.100.075438}
  {\bibfield  {journal} {\bibinfo  {journal} {Phys. Rev. B}\ }\textbf {\bibinfo
  {volume} {100}},\ \bibinfo {pages} {075438} (\bibinfo {year}
  {2019})}\BibitemShut {NoStop}%
\bibitem [{\citenamefont {Jafari}(2019)}]{Jafari2019}%
  \BibitemOpen
  \bibfield  {author} {\bibinfo {author} {\bibfnamefont {S.~A.}\ \bibnamefont
  {Jafari}},\ }\href {\doibase 10.1103/physrevb.100.045144} {\bibfield
  {journal} {\bibinfo  {journal} {Physical Review B}\ }\textbf {\bibinfo
  {volume} {100}} (\bibinfo {year} {2019}),\
  10.1103/physrevb.100.045144}\BibitemShut {NoStop}%
\bibitem [{\citenamefont {Jalali-Mola}\ and\ \citenamefont
  {Jafari}(2019)}]{SaharPolarization}%
  \BibitemOpen
  \bibfield  {author} {\bibinfo {author} {\bibfnamefont {Z.}~\bibnamefont
  {Jalali-Mola}}\ and\ \bibinfo {author} {\bibfnamefont {S.~A.}\ \bibnamefont
  {Jafari}},\ }\href {\doibase 10.1103/physrevb.100.075113} {\bibfield
  {journal} {\bibinfo  {journal} {Physical Review B}\ }\textbf {\bibinfo
  {volume} {100}} (\bibinfo {year} {2019}),\
  10.1103/physrevb.100.075113}\BibitemShut {NoStop}%
\bibitem [{\citenamefont {Farajollahpour}\ \emph {et~al.}(2019)\citenamefont
  {Farajollahpour}, \citenamefont {Faraei},\ and\ \citenamefont
  {Jafari}}]{Tohid2019Spacetime}%
  \BibitemOpen
  \bibfield  {author} {\bibinfo {author} {\bibfnamefont {T.}~\bibnamefont
  {Farajollahpour}}, \bibinfo {author} {\bibfnamefont {Z.}~\bibnamefont
  {Faraei}}, \ and\ \bibinfo {author} {\bibfnamefont {S.~A.}\ \bibnamefont
  {Jafari}},\ }\href {\doibase 10.1103/PhysRevB.99.235150} {\bibfield
  {journal} {\bibinfo  {journal} {Phys. Rev. B}\ }\textbf {\bibinfo {volume}
  {99}},\ \bibinfo {pages} {235150} (\bibinfo {year} {2019})}\BibitemShut
  {NoStop}%
\bibitem [{\citenamefont {Volovik}(2016)}]{Volovik2016}%
  \BibitemOpen
  \bibfield  {author} {\bibinfo {author} {\bibfnamefont {G.~E.}\ \bibnamefont
  {Volovik}},\ }\href {\doibase 10.1134/S0021364016210050} {\bibfield
  {journal} {\bibinfo  {journal} {JETP Letters}\ }\textbf {\bibinfo {volume}
  {104}},\ \bibinfo {pages} {645} (\bibinfo {year} {2016})}\BibitemShut
  {NoStop}%
\bibitem [{\citenamefont {Volovik}(2018)}]{Volovik2018}%
  \BibitemOpen
  \bibfield  {author} {\bibinfo {author} {\bibfnamefont {G.~E.}\ \bibnamefont
  {Volovik}},\ }\href {\doibase 10.3367/ufne.2017.01.038218} {\bibfield
  {journal} {\bibinfo  {journal} {Physics-Uspekhi}\ }\textbf {\bibinfo {volume}
  {61}},\ \bibinfo {pages} {89} (\bibinfo {year} {2018})}\BibitemShut {NoStop}%
\bibitem [{\citenamefont {Nissinen}\ and\ \citenamefont
  {Volovik}(2017)}]{Nissinen2017}%
  \BibitemOpen
  \bibfield  {author} {\bibinfo {author} {\bibfnamefont {J.}~\bibnamefont
  {Nissinen}}\ and\ \bibinfo {author} {\bibfnamefont {G.~E.}\ \bibnamefont
  {Volovik}},\ }\href {\doibase 10.1134/S0021364017070013} {\bibfield
  {journal} {\bibinfo  {journal} {JETP Letters}\ }\textbf {\bibinfo {volume}
  {105}},\ \bibinfo {pages} {447} (\bibinfo {year} {2017})}\BibitemShut
  {NoStop}%
\bibitem [{\citenamefont {Ryder}(2009)}]{RyderGR}%
  \BibitemOpen
  \bibfield  {author} {\bibinfo {author} {\bibfnamefont {L.}~\bibnamefont
  {Ryder}},\ }\href {\doibase 10.1017/cbo9780511809033} {\emph {\bibinfo
  {title} {Introduction to General Relativity}}}\ (\bibinfo  {publisher}
  {Cambridge University Press},\ \bibinfo {year} {2009})\BibitemShut {NoStop}%
\bibitem [{\citenamefont {Schutz}(2009)}]{SchutzGR}%
  \BibitemOpen
  \bibfield  {author} {\bibinfo {author} {\bibfnamefont {B.}~\bibnamefont
  {Schutz}},\ }\href {\doibase 10.1017/cbo9780511984181} {\emph {\bibinfo
  {title} {A First Course in General Relativity}}}\ (\bibinfo  {publisher}
  {Cambridge University Press},\ \bibinfo {year} {2009})\BibitemShut {NoStop}%
\bibitem [{\citenamefont {Mohajerani}\ \emph {et~al.}(2021)\citenamefont
  {Mohajerani}, \citenamefont {Faraei},\ and\ \citenamefont
  {Jafari}}]{Mohajerani2021}%
  \BibitemOpen
  \bibfield  {author} {\bibinfo {author} {\bibfnamefont {A.}~\bibnamefont
  {Mohajerani}}, \bibinfo {author} {\bibfnamefont {Z.}~\bibnamefont {Faraei}},
  \ and\ \bibinfo {author} {\bibfnamefont {S.~A.}\ \bibnamefont {Jafari}},\
  }\href {\doibase 10.1088/1361-648x/abe64e} {\bibfield  {journal} {\bibinfo
  {journal} {Journal of Physics: Condensed Matter}\ }\textbf {\bibinfo {volume}
  {33}},\ \bibinfo {pages} {215603} (\bibinfo {year} {2021})}\BibitemShut
  {NoStop}%
\bibitem [{\citenamefont {Boettcher}\ \emph
  {et~al.}(2020{\natexlab{a}})\citenamefont {Boettcher}, \citenamefont
  {Bienias}, \citenamefont {Belyansky}, \citenamefont {Koll{\'{a}}r},\ and\
  \citenamefont {Gorshkov}}]{Boettcher2020}%
  \BibitemOpen
  \bibfield  {author} {\bibinfo {author} {\bibfnamefont {I.}~\bibnamefont
  {Boettcher}}, \bibinfo {author} {\bibfnamefont {P.}~\bibnamefont {Bienias}},
  \bibinfo {author} {\bibfnamefont {R.}~\bibnamefont {Belyansky}}, \bibinfo
  {author} {\bibfnamefont {A.~J.}\ \bibnamefont {Koll{\'{a}}r}}, \ and\
  \bibinfo {author} {\bibfnamefont {A.~V.}\ \bibnamefont {Gorshkov}},\ }\href
  {\doibase 10.1103/physreva.102.032208} {\bibfield  {journal} {\bibinfo
  {journal} {Physical Review A}\ }\textbf {\bibinfo {volume} {102}} (\bibinfo
  {year} {2020}{\natexlab{a}}),\ 10.1103/physreva.102.032208}\BibitemShut
  {NoStop}%
\bibitem [{\citenamefont {Baek}\ \emph {et~al.}(2009)\citenamefont {Baek},
  \citenamefont {Minnhagen},\ and\ \citenamefont {Kim}}]{Baek2009}%
  \BibitemOpen
  \bibfield  {author} {\bibinfo {author} {\bibfnamefont {S.~K.}\ \bibnamefont
  {Baek}}, \bibinfo {author} {\bibfnamefont {P.}~\bibnamefont {Minnhagen}}, \
  and\ \bibinfo {author} {\bibfnamefont {B.~J.}\ \bibnamefont {Kim}},\ }\href
  {\doibase 10.1103/physreve.79.011124} {\bibfield  {journal} {\bibinfo
  {journal} {Physical Review E}\ }\textbf {\bibinfo {volume} {79}} (\bibinfo
  {year} {2009}),\ 10.1103/physreve.79.011124}\BibitemShut {NoStop}%
\bibitem [{\citenamefont {Koll{\'{a}}r}\ \emph {et~al.}(2019)\citenamefont
  {Koll{\'{a}}r}, \citenamefont {Fitzpatrick},\ and\ \citenamefont
  {Houck}}]{Kollar2019}%
  \BibitemOpen
  \bibfield  {author} {\bibinfo {author} {\bibfnamefont {A.~J.}\ \bibnamefont
  {Koll{\'{a}}r}}, \bibinfo {author} {\bibfnamefont {M.}~\bibnamefont
  {Fitzpatrick}}, \ and\ \bibinfo {author} {\bibfnamefont {A.~A.}\ \bibnamefont
  {Houck}},\ }\href {\doibase 10.1038/s41586-019-1348-3} {\bibfield  {journal}
  {\bibinfo  {journal} {Nature}\ }\textbf {\bibinfo {volume} {571}},\ \bibinfo
  {pages} {45} (\bibinfo {year} {2019})}\BibitemShut {NoStop}%
\bibitem [{\citenamefont {Yekta}\ \emph {et~al.}(2021)\citenamefont {Yekta},
  \citenamefont {Hadipour},\ and\ \citenamefont {Jafari}}]{Yasin2021}%
  \BibitemOpen
  \bibfield  {author} {\bibinfo {author} {\bibfnamefont {Y.}~\bibnamefont
  {Yekta}}, \bibinfo {author} {\bibfnamefont {H.}~\bibnamefont {Hadipour}}, \
  and\ \bibinfo {author} {\bibfnamefont {S.~A.}\ \bibnamefont {Jafari}},\
  }\href@noop {} {\enquote {\bibinfo {title} {How to tune the tilt of a dirac
  cone by atomic manipulations?}}\ } (\bibinfo {year} {2021}),\ \Eprint
  {http://arxiv.org/abs/arXiv:2108.08183} {arXiv:2108.08183} \BibitemShut
  {NoStop}%
\bibitem [{\citenamefont {Kardar}(2007)}]{Kardar2007}%
  \BibitemOpen
  \bibfield  {author} {\bibinfo {author} {\bibfnamefont {M.}~\bibnamefont
  {Kardar}},\ }\href {\doibase 10.1017/cbo9780511815881} {\emph {\bibinfo
  {title} {Statistical Physics of Fields}}}\ (\bibinfo  {publisher} {Cambridge
  University Press},\ \bibinfo {year} {2007})\BibitemShut {NoStop}%
\bibitem [{\citenamefont {Kadanoff}(2000)}]{Kadanoff2000}%
  \BibitemOpen
  \bibfield  {author} {\bibinfo {author} {\bibfnamefont {L.~P.}\ \bibnamefont
  {Kadanoff}},\ }\href {\doibase 10.1142/4016} {\emph {\bibinfo {title}
  {Statistical Physics}}}\ (\bibinfo  {publisher} {{WORLD} {SCIENTIFIC}},\
  \bibinfo {year} {2000})\BibitemShut {NoStop}%
\bibitem [{\citenamefont {Vozmediano}\ \emph {et~al.}(2010)\citenamefont
  {Vozmediano}, \citenamefont {Katsnelson},\ and\ \citenamefont
  {Guinea}}]{Vozmediano2010}%
  \BibitemOpen
  \bibfield  {author} {\bibinfo {author} {\bibfnamefont {M.}~\bibnamefont
  {Vozmediano}}, \bibinfo {author} {\bibfnamefont {M.}~\bibnamefont
  {Katsnelson}}, \ and\ \bibinfo {author} {\bibfnamefont {F.}~\bibnamefont
  {Guinea}},\ }\href {\doibase 10.1016/j.physrep.2010.07.003} {\bibfield
  {journal} {\bibinfo  {journal} {Physics Reports}\ }\textbf {\bibinfo {volume}
  {496}},\ \bibinfo {pages} {109} (\bibinfo {year} {2010})}\BibitemShut
  {NoStop}%
\bibitem [{\citenamefont {Economou}(2006)}]{economou_2006}%
  \BibitemOpen
  \bibfield  {author} {\bibinfo {author} {\bibfnamefont {E.~N.}\ \bibnamefont
  {Economou}},\ }\href@noop {} {\emph {\bibinfo {title} {Greens Functions in
  Quantum Physics}}}\ (\bibinfo  {publisher} {Springer-Verlag Berlin
  Heidelberg},\ \bibinfo {year} {2006})\BibitemShut {NoStop}%
\bibitem [{\citenamefont {Cserti}\ \emph {et~al.}(2011)\citenamefont {Cserti},
  \citenamefont {Sz{\'{e}}chenyi},\ and\ \citenamefont
  {D{\'{a}}vid}}]{Cserti2011}%
  \BibitemOpen
  \bibfield  {author} {\bibinfo {author} {\bibfnamefont {J.}~\bibnamefont
  {Cserti}}, \bibinfo {author} {\bibfnamefont {G.}~\bibnamefont
  {Sz{\'{e}}chenyi}}, \ and\ \bibinfo {author} {\bibfnamefont {G.}~\bibnamefont
  {D{\'{a}}vid}},\ }\href {\doibase 10.1088/1751-8113/44/21/215201} {\bibfield
  {journal} {\bibinfo  {journal} {Journal of Physics A: Mathematical and
  Theoretical}\ }\textbf {\bibinfo {volume} {44}},\ \bibinfo {pages} {215201}
  (\bibinfo {year} {2011})}\BibitemShut {NoStop}%
\bibitem [{Iba(2009)}]{Ibach2009}%
  \BibitemOpen
  \href {\doibase 10.1007/978-3-540-85795-2} {\emph {\bibinfo {title}
  {Festk\"{o}rperphysik}}}\ (\bibinfo  {publisher} {Springer Berlin
  Heidelberg},\ \bibinfo {year} {2009})\BibitemShut {NoStop}%
\bibitem [{\citenamefont {Boettcher}\ \emph
  {et~al.}(2020{\natexlab{b}})\citenamefont {Boettcher}, \citenamefont
  {Bienias}, \citenamefont {Belyansky}, \citenamefont {Koll{\'{a}}r},\ and\
  \citenamefont {Gorshkov}}]{Gorshkov2020}%
  \BibitemOpen
  \bibfield  {author} {\bibinfo {author} {\bibfnamefont {I.}~\bibnamefont
  {Boettcher}}, \bibinfo {author} {\bibfnamefont {P.}~\bibnamefont {Bienias}},
  \bibinfo {author} {\bibfnamefont {R.}~\bibnamefont {Belyansky}}, \bibinfo
  {author} {\bibfnamefont {A.~J.}\ \bibnamefont {Koll{\'{a}}r}}, \ and\
  \bibinfo {author} {\bibfnamefont {A.~V.}\ \bibnamefont {Gorshkov}},\ }\href
  {\doibase 10.1103/physreva.102.032208} {\bibfield  {journal} {\bibinfo
  {journal} {Physical Review A}\ }\textbf {\bibinfo {volume} {102}} (\bibinfo
  {year} {2020}{\natexlab{b}}),\ 10.1103/physreva.102.032208}\BibitemShut
  {NoStop}%
\bibitem [{\citenamefont {Maciejko}\ and\ \citenamefont
  {Rayan}(2021)}]{Maciejko2021}%
  \BibitemOpen
  \bibfield  {author} {\bibinfo {author} {\bibfnamefont {J.}~\bibnamefont
  {Maciejko}}\ and\ \bibinfo {author} {\bibfnamefont {S.}~\bibnamefont
  {Rayan}},\ }\href {\doibase 10.1126/sciadv.abe9170} {\bibfield  {journal}
  {\bibinfo  {journal} {Science Advances}\ }\textbf {\bibinfo {volume} {7}}
  (\bibinfo {year} {2021}),\ 10.1126/sciadv.abe9170}\BibitemShut {NoStop}%
\bibitem [{\citenamefont {Campi}\ and\ \citenamefont
  {Bianconi}(2015)}]{Campi2015}%
  \BibitemOpen
  \bibfield  {author} {\bibinfo {author} {\bibfnamefont {G.}~\bibnamefont
  {Campi}}\ and\ \bibinfo {author} {\bibfnamefont {A.}~\bibnamefont
  {Bianconi}},\ }\href {\doibase 10.1007/s10948-015-3326-9} {\bibfield
  {journal} {\bibinfo  {journal} {Journal of Superconductivity and Novel
  Magnetism}\ }\textbf {\bibinfo {volume} {29}},\ \bibinfo {pages} {627}
  (\bibinfo {year} {2015})}\BibitemShut {NoStop}%
\bibitem [{\citenamefont {Farajollahpour}\ and\ \citenamefont
  {Jafari}(2020)}]{Tohid2020Synthetic}%
  \BibitemOpen
  \bibfield  {author} {\bibinfo {author} {\bibfnamefont {T.}~\bibnamefont
  {Farajollahpour}}\ and\ \bibinfo {author} {\bibfnamefont {S.~A.}\
  \bibnamefont {Jafari}},\ }\href {\doibase 10.1103/physrevresearch.2.023410}
  {\bibfield  {journal} {\bibinfo  {journal} {Physical Review Research}\
  }\textbf {\bibinfo {volume} {2}} (\bibinfo {year} {2020}),\
  10.1103/physrevresearch.2.023410}\BibitemShut {NoStop}%
\end{thebibliography}%

\newpage
\pagebreak
\renewcommand{\thesection}{S\arabic{section}}  
\renewcommand{\thetable}{S\arabic{table}}  
\renewcommand{\thefigure}{S\arabic{figure}}
\renewcommand{\theequation}{S\arabic{equation}}

\setcounter{equation}{0}
\setcounter{page}{1}
\setcounter{figure}{0}
\setcounter{section}{0}

\begin{center}
{\bf Supplementary material:  
The circuit realization of tilted Dirac cone: A platform for fabrication of curved spacetime geometry on a chip
}
\end{center}
In this annex we provide some calculational details related to the main text. 

\subsection{Conditions for Dirac Crossing}

The zeros of the $\Delta(\bs k)$ given in Eq.~\eqref{absdelta.eqn} are obtained as follows:
One can view it as a quadratic equation in $\cos (k_y/2)$ and find the following roots,
\begin{equation}
\cos \frac{k_y}{2}=\frac{-(1+c)\cos \frac{k_x\sqrt{3}}{2} \pm |1-c|\sqrt{\cos^2 \frac{k_x\sqrt{3}}{2}-1}}{2}.\nn
\label{cosquad.eqn}
\end{equation}
Let us begin the analysis of this equation by first looking at a limit in case $c=1$.
In this situation the coefficient of the square root vanishes and the two roots 
degenerate into the simple equation $\cos (k_y/2) = -\cos (k_x\sqrt{3}/2)$
or equivalently  $k_y = 2\pi \pm k_x\sqrt{3}$, 
which represents two straight lines in k-space inside Brillouin zone (BZ). 
For the generic case, $c\neq1$ the expression under the square root is most of the time 
negative (since cosine is always bounded by $-1$ and $+1$), and the only chance to avoide complex
frequencies is to set the square root to zero, namely $\cos (k_x\sqrt{3}/2) = \pm 1 $ which implies
$\cos (k_y/2) = \mp (1+c)/2 $. 
In order to have real solution for $k_y$, the (already positive) parameter $c$ must be smaller than one. 
Therefore the condition to have zero band gap is $c\leq 1$ where the the limiting case $c=1$ marks the transition between zero and non-zero band gap. 
This explains why in the main text we have restricted ourselves to the region $0 < c < 1$. 

\subsection{Band exterma}
As pointed out in the main text, the band exterma are obtained from $\Nab_{\bs k}\Omega=0$. 
In addition to the van-Hove singularities at $\mathbf{M}_1$ and $\mathbf{M}_2$, depending on the parameter values,
one can also have two other exterma $\mathbf{S}_1$ and $\mathbf{S}_2$ listed in Tab.~\ref{table1}. But they may not always exist, 
because the arccosine argument is restricted to lie between $\pm1$. The value of $\Omega$ at $\mathbf{S}_1$ is greater than the level $\bar\omega_0^2$ of the Dirac point. 
The value of $\Omega$ at $\mathbf{S}_2$ is greater than $\Omega_+(\mathbf{\Gamma})$ and the level of $\Omega$ at $\mathbf{S}_1$ is greater than $\Omega$ at Dirac point. 
The exterma at  $\mathbf{S}_1$ and $\mathbf{S}_2$ do not contribute a divergent DOS, and hence do not alter
the impedance spectroscopy. We discuss them here for the mathematical completeness of our theory.
In fact they are maxima or minima with no saddle character that contribute a constant jump into the DOS. 
Fig.~\ref{fig2} of the main text has been produced for parameter ranges where the above points do not exist.

\begin{figure}[t]
\centering
\includegraphics[clip, trim=7.6cm 18.7cm 7.6cm 4.6cm, width=0.46\textwidth]{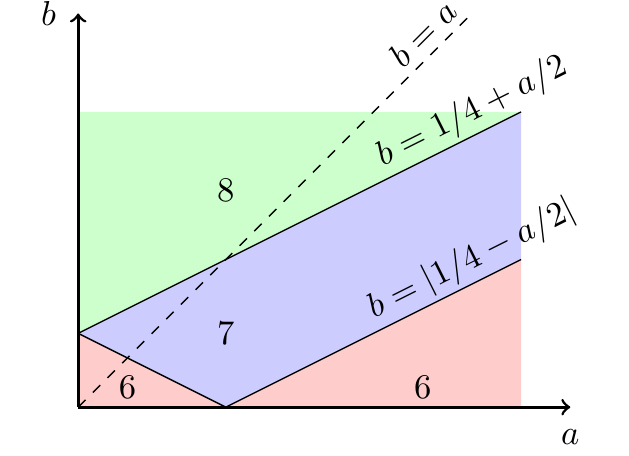}
\caption{The number of exterma in the BZ for various regions of the $ab$ plane. 
On the special line $b=a$, the value of $\Omega$ for $\mathbf{M}_1$ and $\mathbf{M}_2$ points are the same, 
therefore on this dashed line the numeber of exterma reduce by two.}
\label{figSM}
\end{figure}

Since the number of exterma varies in the parameter space, 
it is appropriate to specify the number of exterma in the plane of parameter $(a,b)$. 
The two exterma $\mathbf{S}_1$ and $\mathbf{S}_2$  exists when  $1/4+a/2<b$ and $-b<1/4-a/2<b$ (or equivalently $|1/4-a/2|<b$) are satisfied.
The intersection of the above conditions gives the green region in Fig.~\ref{figSM} in $ab$ plane that has the maximmal number $8$ exterma. 
Outside these region denoted by pink, the above two exterma do not exist and hence we have $6$ exterma. In the purple
region where only  the condition for the existence of $\mathbf{S}_2$ is satisfied we have $7$ exterma. The line
$a=b$ passes through all three regions. When this condition is satisfied, the saddle points 
$\mathbf{M}_1$ and $\mathbf{M}_2$ coincide and the corresponiding number of exterma on this line 
is reduced by $2$. The data for Fig.~\ref{fig2} and Fig.~\ref{fig3} are produced in the pink range of the parameter space. 

\subsection{Details of numeric simulations}
Requringt the inductance parameters $a$, $b$ and $c$ to be positive, the allowed range for  the parameter $\alpha_y$ 
(that determines the tilt $\zeta_y$) becomes
\begin{equation}
-\frac{(1+c)(\bar\omega_0^2-3-c)}{\sqrt{(3+c)(1-c)}}<\alpha_y<\frac{1}{2}(\bar\omega_0^2-3-c)\sqrt{\frac{1-c}{3+c}}.
\end{equation}
As it can be seen, non zero values for $\alpha_y$ can be reached subject to the the condition $\bar\omega_0^2>3+c$. 
Since we want to have dirac points, we keep $0\leq c <1$. In our numeric simulations we fix the level of Dirac points by $\bar\omega_0^2=4$. 
We impose additional constraint on $\zeta$ in order to stay in the pink regions of Fig.~\ref{fig2}.
The method we use to compute the DOS is to simply count the numeber of $\bs k$ points in k-space at each value of $\Omega$. 

As we noted below Eq.~\eqref{imp.eqn}, all the exterma of the DOS are contained in the sublattice-diagonal impendance (with arbitrary unit cell position). Therefore, we calculate, $\operatorname{Re}\{Z_{AA}(\bs{a}_1)\}$, the impedance between sublattice A of two sites at distance $\bs{a}_1$ from each other
where $\bs a_1$ is the basis vector of the underlying honecomb lattice. 

In Fig.~\ref{fig2} the curves of impedance and DOS vs. $\Omega$ are shown. Note how the van Hove singularities of both curves coincide, as expected. 
The value of $\Omega$ at the van Hove singularities are exactly those reported in Tab.~\ref{table1} for the DOS. 
Near van hove logarithmic singularities, which are at $\mathbf{M}_1$ and $\mathbf{M}_2$ in k-space, the real part of impedance diverges 
that can be interpreted as "open circuit" situation for resistance. 
The Dirac point intself is manifested as zero impedance point that is reminiscent of "short circuit" situation for resistance.

\subsection{Tilt parameter as slope of the impedance lineshape}
The value of the tilt parameter can be directly extracted from the linear part of the DOS that coincides with the local impedance
at separation $\bs r=\bs a_1$. 
The constant frequency surface near the Dirac crossing level is an ellipsoid, therefore the DOS can be simply obtained by computing the $\Omega$ derivative of ellipsoid area which gives,
\begin{equation}
\rho(\Omega)\approx V_0\frac{(3+c)|\Omega-\bar\Omega_0|}{32\pi\sqrt{3}\bar\omega_0^3v_y^3(1-\zeta_y^2)^{3/2}}
\end{equation}     
where $\zeta_y=v_{ty}/v_y$ is the tilt along $y$ axis. The "redshift factor" $1/\sqrt{1-\zeta^2}$ is the characteristic of
many spectroscopic quantities~\cite{Mohajerani2021} and is a hallmark of the peculiar spacetime structure encoded in the
metric~\eqref{metric1.eqn} of the main text. The same quantity connects the impedance of non-tilted and tilted Dirac cone circuits.

\end{document}